\documentclass{iau379}

\usepackage{amsmath}
\usepackage{multirow}
\usepackage{graphicx,float}
\usepackage{hyperref}

\begin{document}

\lefttitle{Sakowska et al.}
\righttitle{SFH of the SMC: the shell overdensity}

\jnlPage{1}{7}
\jnlDoiYr{2023}
\doival{10.1017/xxxxx}

\aopheadtitle{Proceedings of IAU Symposium 379}
\editors{P. Bonifacio,  M.-R. Cioni, F. Hammer, M. Pawlowski, and S. Taibi, eds.}

\title{Star Formation History of the Small Magellanic Cloud: the shell substructure}

\author{Sakowska J. D.$^1$, No\"{e}l N. E. D.$^1$, Ruiz-Lara T.$^2$, and Gallart C.$^{3,4}$}
\affiliation{$^1$ Department of Physics, University of Surrey, Guildford, GU2 7JH, United Kingdom \\
$^2$ Universidad de Granada (ugr), Departamento de F\'{i}sica Te\'{o}rica y del Cosmos, Campus Fuente Nueva, Edificio Mecenas,
18071 Granada, Spain \\
$^3$ Instituto de Astrof\'{i}sica de Canarias, V\'{i}a L\'{a}ctea s/n, E-38205 La Laguna, Tenerife, Spain \\
$^4$ Departamento de Astrof\'{i}sica, Universidad de La Laguna, E-38200 La Laguna, Tenerife, Spain}

\begin{abstract}
We present the spatially resolved star formation history (SFH) of a shell-like structure located in the northeastern Small Magellanic Cloud (SMC). We quantitatively obtain the SFH using unprecedented deep photometric data ($g \sim$ 24 magnitude) from the SMASH survey and colour-magnitude diagram (CMD) fitting techniques. We consider, for the first time, the SMC's line-of-sight depth and its optical effects on the CMDs. The SFH presents higher accuracy when a line-of-sight depth of $\sim$3 Kpc is simulated.

We find young star formation enhancements at $\sim$150 Myr, $\sim$200 Myr, $\sim$450 Myr, $\sim$650 Myr, and $\sim$1 Gyr. Comparing the shell's SFH with the Large Magellanic Cloud's (LMC) northern arm SFH we show strong evidence of synchronicity from at least the past $\sim$2.8 Gyr and, possibly, the past $\sim$3.5 Gyr. Our results place constraints on the orbital history of the Magellanic Clouds which, potentially, have implications on their dynamical mass estimates.

\end{abstract}


\maketitle

\section{Introduction}

\begin{figure}
\centering
    \includegraphics[scale=.28]{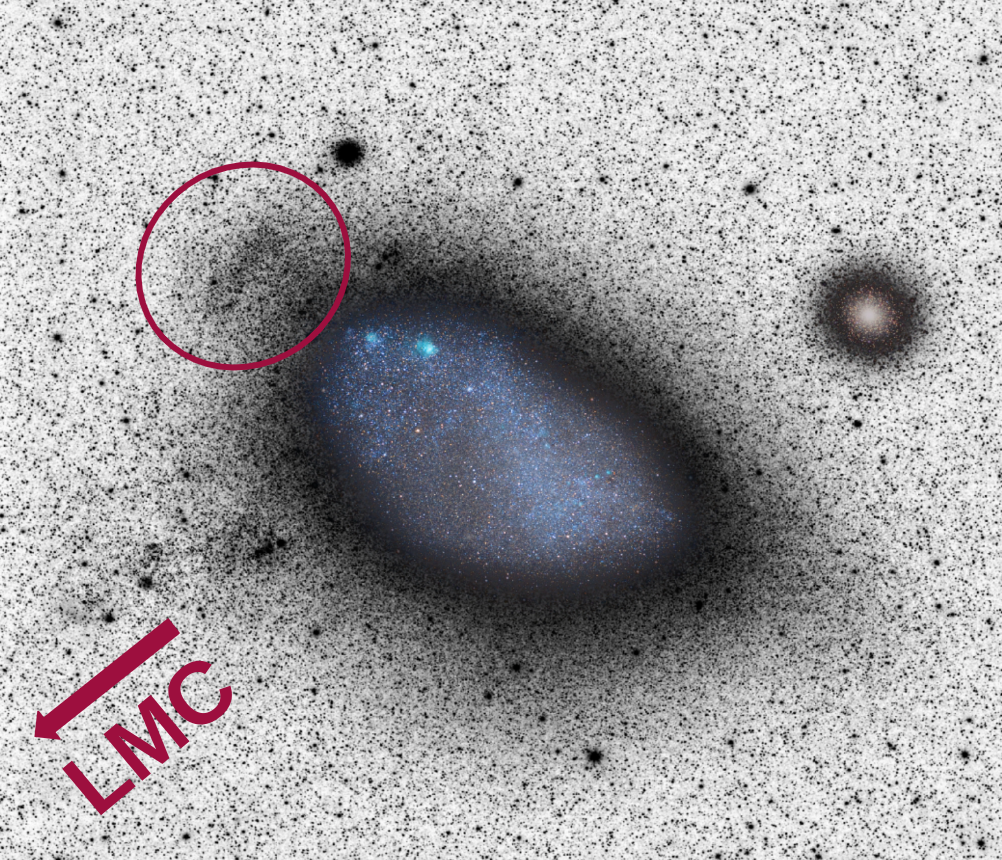}
    \includegraphics[scale=.29]{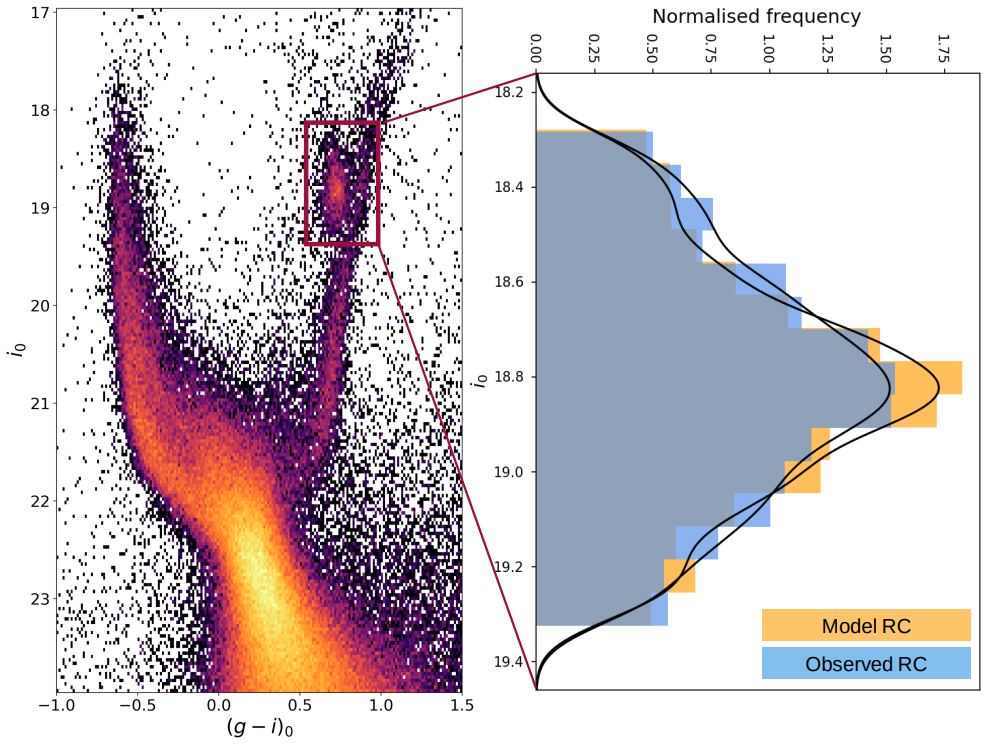}
    \caption{Top panel: SMC image taken from \cite{MD2019} (top panel). Bottom panel: CMD of the shell (left) and the  observed vs. model luminosity function of the red clump (right). The model RC shown here is taken from our first SFH (see \hyperref[sec2]{Section 2}).}
    \label{SMC}
\end{figure}

The faint peripheries of galaxies contain stellar fossil records of a system's past events and, hence, hold the keys to understanding their formation and evolution (e.g. \citealt{ElmegreenHunter2017}). The Small and Large Magellanic Cloud (SMC/LMC), located just $\sim$50 Kpc and $\sim$60 Kpc away from us (\citealt{Pietrzynski2019, Graczyk2020}), offer an outstanding natural laboratory to study the resolved stellar populations of two tidally disrupting galaxies in exquisite detail. Given the SMC's smaller total mass - estimated at $\sim$1/10th of the LMC (De Leo et al., submitted to MNRAS) - it is likely the LMC's gravitational influence played a key role in tidally shaping the SMC (e.g. \citealt{deleo2020}).
In \hyperref[SMC]{Fig. 1} (top) we see the SMC with a plethora of stellar substructures around its main body, likely to have been produced during tidal interactions with the LMC. What we cannot see, however, are elongations along the line-of-sight which, at the extremes, could potentially be almost half of the distance to the SMC from us ($\sim$20 - 30 Kpc in the east, e.g. \citealt{Hatzidimitriou1989, Nidever2013, Scowcroft2016, Ripepi2017, Tatton2021}). 


In this work, we focus on a shell-like overdensity of stars in the outskirts of the SMC, towards its north-east direction \citep{MD2019}  seen in   \hyperref[SMC]{Fig. 1} (top panel). We derive a spatially resolved star formation history (SFH) of the shell using deep colour magnitude diagrams (CMDs) from SMASH data (\citealt{Nidever2017,Nidever2021}; ugriz $\sim$24th mag). To achieve a quantitative determination of the SFH we have taken into account, for the first time, the puzzling line-of-sight depth of the SMC using the red clump (RC). 
In \hyperref[sec2]{Section 2} we describe the analysis, including the derivation of the SFH. In \hyperref[sec3]{Section 3} we discuss our results. We present the conclusions in \hyperref[sec4]{Section 4}. 

\section{Star formation history derivation}\label{sec2}

The bottom panel in
\hyperref[SMC]{Fig. 1} (left) shows the CMD of the SMC's shell overdensity. While the line-of-sight depth affects the entire CMD, this effect can be seen more conspicuously in the  RC region (i.e. at $\sim 0.5 \le (g-i)_{0} \le 1.0, 18.2 \le i_{0} \le 19.5 $). \hyperref[SMC]{Fig. 1} (right) shows the observed RC luminosity function (blue) against a theoretical RC luminosity function for the SMC region (see caption for details). The line-of-sight depth in the shell region causes the observed RC to be more elongated in magnitude than the RC resulted from the models. We will discuss this in greater detail below.  

To derive the SFH we simulate the observed distribution of stars into a `solution' CMD by quantitatively comparing the observed and synthetic `model' CMD. 
First, we quantified the photometric completeness of the observed CMD (affected by stellar crowding, blending, and other potential errors) by performing artificial star tests (ASTs) following the procedure described in \cite{Rusakov2021}. For the CMD fitting, we only considered high-quality photometry above the 50\% completeness threshold, which for the shell lies well below the main-sequence turn off at $M_{g}\sim 22$ mag for the SMC.

We generated the shell's model CMD using the solar-scaled BaSTI-IAC stellar evolution models (\citealt{Hidalgo2018, Pietrinferni2004}), with 2022 updates (S. Cassisi, private communication), containing $1.5 \times 10^{8}$  stars with uniform distributions in age ($0.03 \leq$  age $\leq 14$ Gyr) and metallicity ($0.00001 \leq Z \leq 0.025$). 
We then simulated  observational effects on the model CMD  using \verb|DisPar| (see Sakowska et al., in prep.).  \verb|DisPar| used the results from the ASTs to `disperse' the stars from their actual positions on the model CMD in order to simulate the aforementioned observational effects seen in the observed CMD.

To derive the first version of the SFH, we used a Poisson adapted $\chi^{2}$ minimisation algorithm (\citealt{Bernard2015, Bernard2018}) that obtained the best combination of single stellar populations (SSP) from a synthetic CMD that fits the distribution of stars in the observed CMD.  The best fitting combinations were reproduced in a `solution' CMD from which we obtained the SFH.



\begin{figure}
\centering
    \includegraphics[scale=.3]{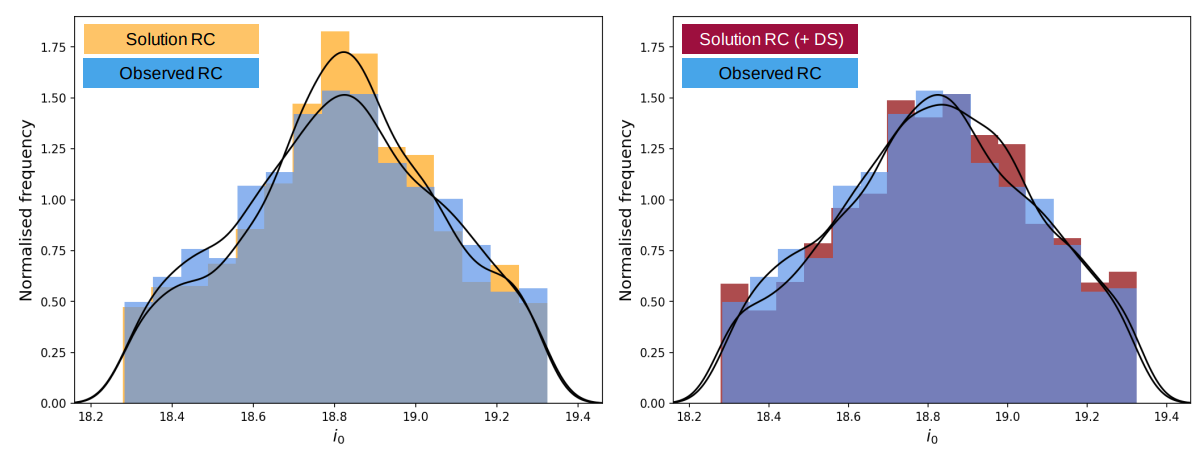}
    \caption{Finding the distance spread which must be applied to the solution red clump (left, orange) to best reproduce the observed red clump (left, blue) for the shell. We found a gaussian-like distance spread of 3 Kpc best reproduced the observed luminosity function (right, burgundy).}\label{ShellRChists}

\end{figure}

In order to improve our SFH derivation, we must quantify the distance spread effect (DS) seen in our observed CMD and simulate it in the model CMD we fit it against.  To do so, we considered the morphology of the red clump region in the CMD (see \hyperref[SMC]{Fig. 1} and description above).
We illustrate this procedure in \hyperref[ShellRChists]{Fig. 2}: the left panel shows the the observed (blue) and the solution (orange) RC luminosity functions and the right panel shows the distance-spread distribution which best reproduced the observed RC luminosity function (burgundy). We obtain the most accurate solution when simulating a DS of $\sim$3 Kpc.



\section{Results and discussion}\label{sec3}

\begin{figure}
\centering
    \includegraphics[scale=.4]{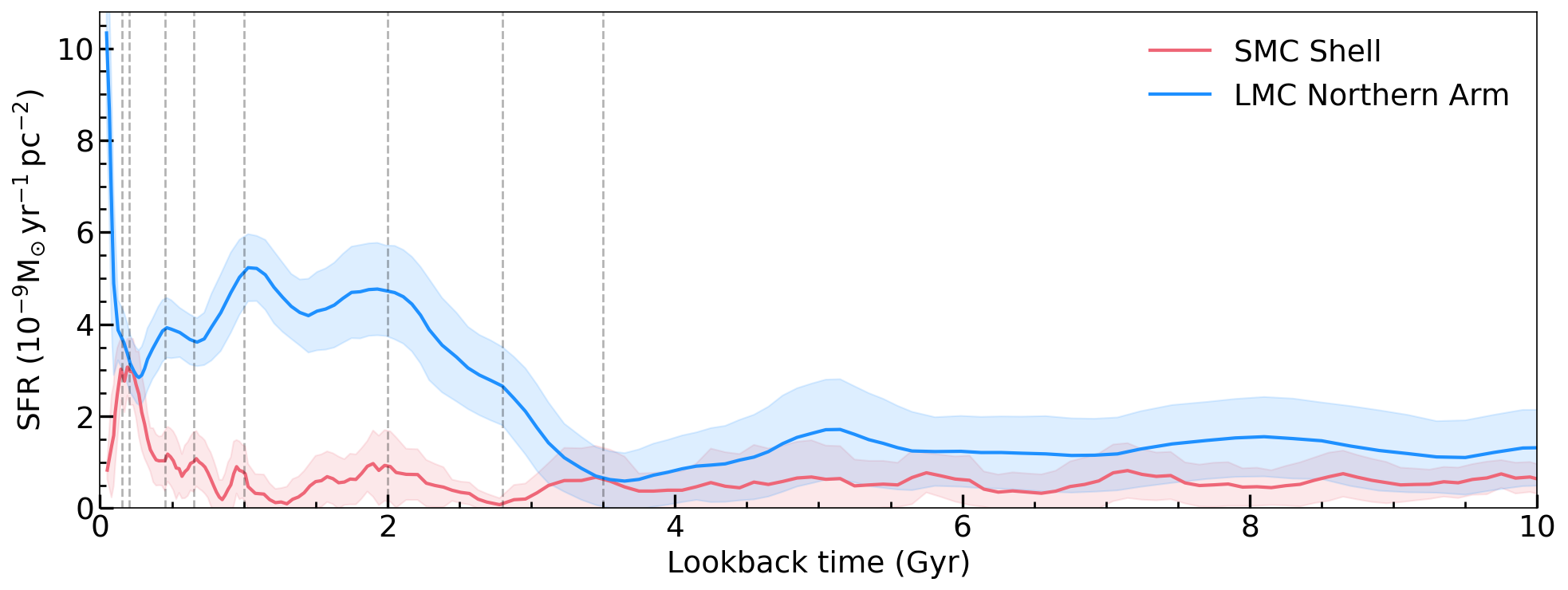}
    \caption{Star formation history of the shell with its distance spread accounted for (pink). Overlayed is the SFH of the LMC's northern arm also from SMASH data (blue, \citealt{Ruiz-Lara2020}). We highlight SFR synchronicity at $\sim$2.8 Gyr, $\sim$2 Gyr, $\sim$1 Gyr, $\sim$450 Myr, $\sim$200 Myr, $\sim$150 Myr and the shell's SFR peak at $\sim$650 Myr. Note the matching, potentially synchronous SFR at $\sim$3.5 Gyr.}
    \label{SFH}
\end{figure}

In \hyperref[SFH]{Fig. 3} we present the SFH of the shell overdensity (pink) considering the line-of-sight depth effect (with a DS of $\sim3$ Kpc). We also included the SFH of the LMC's northern arm (\citealt{Ruiz-Lara2020}) in \hyperref[SFH]{Fig. 3} (blue). The comparison is possible given that \cite{Ruiz-Lara2020} follow the same methodology we presented here but without including the line-of-sight depth since this effect is negligible in the LMC. As seen in \hyperref[SFH]{Fig. 3}, the SFH of the shell presents clear enhancements at young ages of $\sim$150 Myr, $\sim$200 Myr, $\sim$450 Myr, $\sim$650 Myr and $\sim$1 Gyr, implying a possible synchronicity with the LMC's northern arm. This agrees with the results from \cite{Massana2022} who also followed the same methodology. Although the synchronicity in the SFHs is clear for at least the  past $\sim$2.8 Gyr,   the shell and the LMC's northern arm SFHs seem to be very similar in the past $\sim$3.5 Gyr,  indicating a potentially earlier start to the synchronicity.  

\section{Conclusions}\label{sec4}

We present the first quantitative SFH of the SMC's shell overdensity accounting for line-of-sight depth effects. We found the best CMD fit to occur when simulating a line-of-sight depth of $\sim$3 Kpc. The SFH presented here shows evidence for SMC/LMC co-evolution dating at least $\sim$2.8 Gyr, but potentially even from $\sim$3.5 Gyr ago, hinting that the LMC played a key role in tidally shaping the SMC's shell substructure.


\begin{thebibliography}{}


\bibitem[\protect\citeauthoryear{Bernard et al.}{2015}]{Bernard2015} Bernard E.~J., Ferguson A.~M.~N., Chapman S.~C., et al., 2015, MNRAS, 453, L113

\bibitem[\protect\citeauthoryear{Bernard et al.}{2018}]{Bernard2018} Bernard E.~J., Schultheis M., Di Matteo P., et al., 2018, MNRAS, 477, 3507

\bibitem[\protect\citeauthoryear{De Leo et al.}{2020}]{deleo2020} De Leo M., Carrera R., No{\"e}l N.~E.~D., et al., 2020, MNRAS, 495, 98


\bibitem[\protect\citeauthoryear{Elmegreen \& Hunter}{2017}]{ElmegreenHunter2017} Elmegreen B.~G., Hunter D.~A., 2017, ASSL, 434, 115

\bibitem[\protect\citeauthoryear{Graczyk et al.}{2020}]{Graczyk2020} Graczyk D., Pietrzy{\'n}ski G., Thompson I.~B., et al., 2020, ApJ, 904, 13

\bibitem[\protect\citeauthoryear{Hatzidimitriou \& Hawkins}{1989}]{Hatzidimitriou1989} Hatzidimitriou D., Hawkins M.~R.~S., 1989, MNRAS, 241, 667

\bibitem[\protect\citeauthoryear{Hidalgo et al.}{2018}]{Hidalgo2018} Hidalgo S.~L., Pietrinferni A., Cassisi S., et al., 2018, ApJ, 856, 125

\bibitem[\protect\citeauthoryear{Mart{\'\i}nez-Delgado et al.}{2019}]{MD2019} Mart{\'\i}nez-Delgado D., Vivas A.~K., Grebel E.~K., et al., 2019, A\&A, 631, A98

\bibitem[\protect\citeauthoryear{Massana et al.}{2022}]{Massana2022} Massana P., Ruiz-Lara T., No{\"e}l N.~E.~D., et al., 2022, MNRAS, 513, L40

\bibitem[\protect\citeauthoryear{Nidever et al.}{2013}]{Nidever2013} Nidever D.~L., Monachesi A., Bell E.~F., et al., 2013, ApJ, 779, 145

\bibitem[\protect\citeauthoryear{Nidever et al.}{2017}]{Nidever2017} Nidever D.~L., Olsen K., Walker A.~R., et al., 2017, AJ, 154, 199

\bibitem[\protect\citeauthoryear{Nidever et al.}{2021}]{Nidever2021} Nidever D.~L., Olsen K., Choi Y., et al., 2021, AJ, 161, 74

\bibitem[\protect\citeauthoryear{Pietrinferni et al.}{2004}]{Pietrinferni2004} Pietrinferni A., Cassisi S., Salaris M., et al., 2004, ApJ, 612, 168



\bibitem[\protect\citeauthoryear{Pietrzy{\'n}ski et al.}{2019}]{Pietrzynski2019} Pietrzy{\'n}ski G., Graczyk D., Gallenne A., et al., 2019, Nature, 567, 200

\bibitem[\protect\citeauthoryear{Ripepi et al.}{2017}]{Ripepi2017} Ripepi V., Cioni M.-R.~L., Moretti M.~I., et al., 2017, MNRAS, 472, 808

\bibitem[\protect\citeauthoryear{Ruiz-Lara et al.}{2020}]{Ruiz-Lara2020} Ruiz-Lara T., Gallart C., Monelli M., et al., 2020, A\&A, 639, L3

\bibitem[\protect\citeauthoryear{Rusakov et al.}{2021}]{Rusakov2021} Rusakov V., Monelli M., Gallart C., et al, 2021, MNRAS, 502, 642

\bibitem[\protect\citeauthoryear{Scowcroft et al.}{2016}]{Scowcroft2016} Scowcroft V., Freedman W.~L., Madore B.~F., et al., 2016, ApJ, 816, 49

\bibitem[\protect\citeauthoryear{Tatton et al.}{2021}]{Tatton2021} Tatton B.~L., van Loon J.~T., Cioni M.-R.~L.,et al., 2021, MNRAS, 504, 2983
    

\end{thebibliography}
\end{document}